\documentclass[journal]{IEEEtran}
\ifCLASSINFOpdf
\else
\fi

\usepackage[dvipsnames]{xcolor}
\usepackage{cite}
\usepackage[latin1]{inputenc}
\usepackage{amsmath}
\usepackage{amsfonts}
\usepackage{amssymb}
\usepackage{upgreek}
\usepackage{graphicx}
\usepackage{mathtools}
\usepackage{calc}
\usepackage{caption}
\usepackage{subcaption}
\usepackage{balance}
\usepackage{multirow}
\usepackage{hyperref}

\usepackage{booktabs}
\usepackage[linesnumbered,ruled,vlined]{algorithm2e}

\newtheorem{assumption}{Assumption}

\newcommand{\rr}{\mathbb{R}}

\begin{document}
\bstctlcite{IEEEexample:BSTcontrol}

\title{End-to-End Deep Fault Tolerant Control}

\author{Daulet Baimukashev,
        Bexultan Rakhim,
        Matteo Rubagotti,~\IEEEmembership{Senior Member,~IEEE,}\\ and
        Huseyin Atakan Varol,~\IEEEmembership{Senior Member,~IEEE}
        % <-this % stops a space
        
\thanks{D. Baimukashev is with the Institute of Smart Systems and Artificial Intelligence, Nazarbayev University, 010000 Nur-Sultan, Republic of Kazakhstan. Email address: {\tt\small\ daulet.baimukashev@nu.edu.kz.}}% <-this % stops a space
\thanks{B. Rakhim, M. Rubagotti, and H. A. Varol are with the Department of Robotics and Mechatronics, Nazarbayev University, 010000 Nur-Sultan, Republic of Kazakhstan. Email addresses: {\tt\small\{bexultan.rakhim, matteo.rubagotti, ahvarol\}@nu.edu.kz.}}% <-this % stops a space
\thanks{This work was partially supported by the Nazarbayev University Faculty Development Competitive Research Grant no. 240919FD3915.}
\thanks{This work has been accepted for publication on the IEEE/ASME Transactions on Mechatronics. DOI: 10.1109/TMECH.2021.3100150.}
\thanks{\copyright 2021 IEEE. Personal use of this material is permitted. Permission from IEEE must be obtained for all other uses, in any current or future media, including reprinting/republishing this material for advertising or promotional purposes, creating new collective works, for resale or redistribution to servers or lists, or reuse of any copyrighted component of this work in other works.}
}

%%%%%%%%%%%%% The paper headers
\markboth{}%
{Shell \MakeLowercase{\textit{et al.}}: Bare Demo of IEEEtran.cls for IEEE Journals}

\maketitle

\begin{abstract}
PUBLISHED ON IEEE/ASME TRANSACTIONS ON MECHATRONICS, DOI: 10.1109/TMECH.2021.3100150. Ideally, accurate sensor measurements are needed to achieve a good performance in the closed-loop control of mechatronic systems. As a consequence, sensor faults will prevent the system from working correctly, unless a fault-tolerant control (FTC) architecture is adopted. As model-based FTC algorithms for nonlinear systems are often challenging to design, this paper focuses on a new method for FTC in the presence of sensor faults, based on deep learning. The considered approach replaces the phases of fault detection and isolation and controller design with a single recurrent neural network, which has the value of past sensor measurements in a given time window as input, and the current values of the control variables as output. This end-to-end deep FTC method is applied to a mechatronic system composed of a spherical inverted pendulum, whose configuration is changed via reaction wheels, in turn actuated by electric motors. The simulation and experimental results show that the proposed method can handle abrupt faults occurring in link position/velocity sensors. The provided supplementary material includes a video of real-world experiments and the software source code.
\end{abstract}

% Note that keywords are not normally used for peer-review papers.

\IEEEpeerreviewmaketitle

\section{Introduction} %add the image with schematics

\begin{figure}[b!]{}
    \centering
   \includegraphics[width=1\columnwidth]{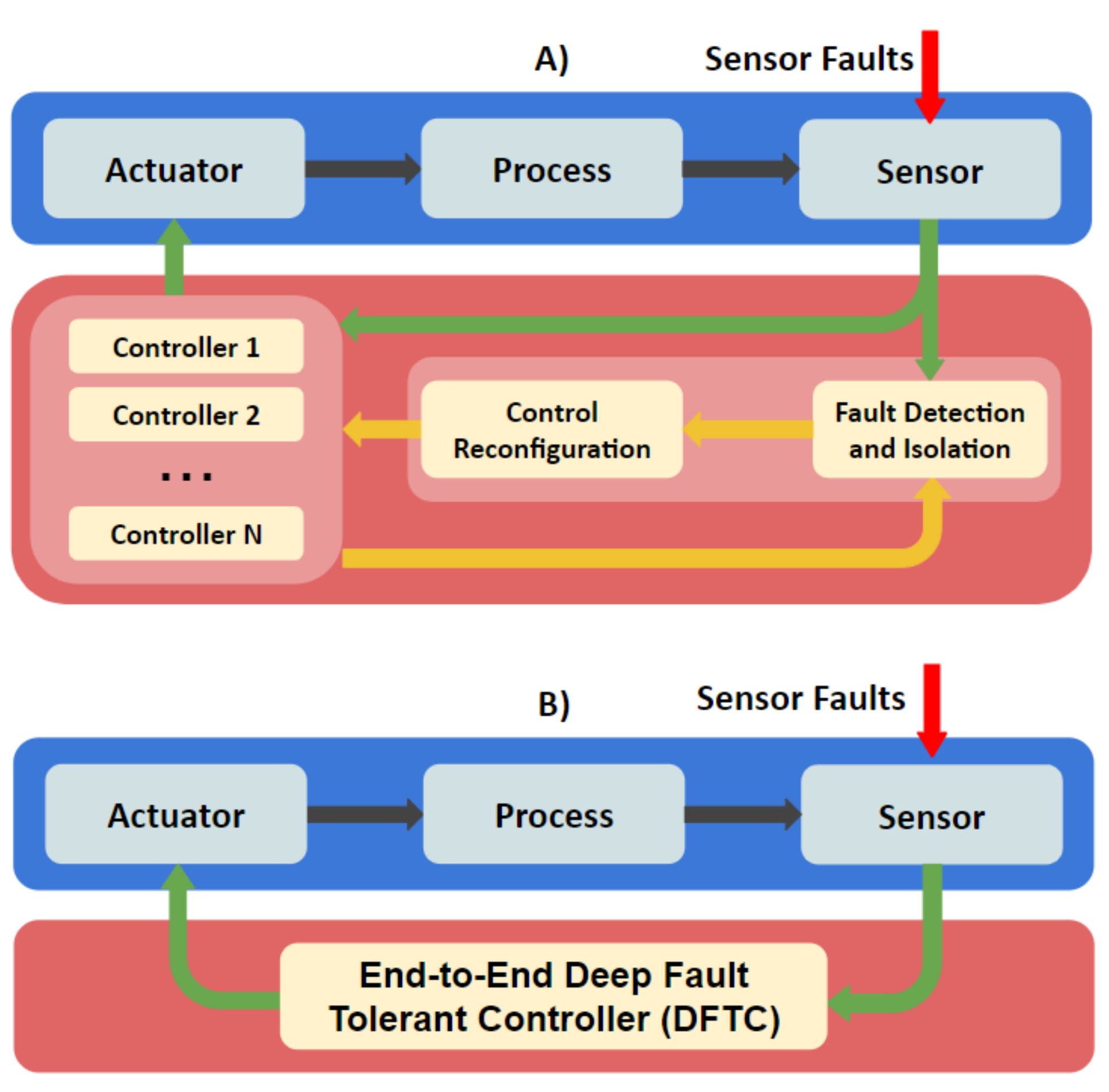}
    \caption{Schematics of the classical FTC system~(A) and neural network-based end-to-end DFTC~(B).}
    \label{fig:schematics}
\end{figure}

\IEEEPARstart{S}{tate}-of-the-art mechatronic systems are operated by advanced control algorithms that allow them to satisfy requirements in terms of performance and safety. However, as the operation of the control system heavily relies on sensor information and on the correct functioning of all system elements, a malfunction in one of these components will likely cause considerable performance degradation or even system instability. This motivation has led to the development of fault-tolerant control (FTC) systems \cite{blanke2006diagnosis}, which can maintain stability and acceptable levels of performance even in the presence of malfunctions. FTC methods are typically divided into passive and active~\cite{benosman2010survey}: in passive methods, the controller is designed to be a-priori robust to a number of expected faults; on the other hand, active methods typically rely on a fault detection and isolation (FDI) module that can effectively address the characteristics of the fault and interact with a control reconfiguration block in real time (see Fig.~\ref{fig:schematics}).

Several methods exist for FDI, from simple limit checking to more advanced strategies, such as those based on parity equations or state observers \cite{isermann2006fault}. Among recent works on FDI-based FTC, researchers have considered the use of higher-order sliding mode observers for fault tolerant speed tracking control of an electric vehicle powered by permanent-magnet synchronous motors~\cite{kommuri2016robust}. Also, in the FTC scheme presented in \cite{manohar2017current}, a third-difference operator was employed in the line current of an induction motor to detect sensor faults, while the actual value of the line current was estimated via a flux-linkage observer. A FTC method based on a motor armature current observer was defined in \cite{li2018sensor}, aimed at guaranteeing acceptable performance in the gear-shifting engaging process of an automated manual transmission, while the authors of \cite{boem2019distributed} combined a distributed FDI algorithm with tube-based model predictive control. Rather than focusing on a specific application, some recent papers have proposed FDI frameworks that are valid for certain classes of systems: for example, sliding mode observers and linear descriptor-based observers have been proposed for fault estimation in Markov jump systems, respectively in \cite{yang2019reduced} and \cite{yang2019descriptor}. The described FDI methods, which will be referred to in the remainder of the paper as \emph{model-based} methods, require the explicit availability of model equations, and in many cases are difficult to design. 

As an alternative to model-based strategies, one can rely on methods based on machine learning, which will be referred to as \emph{data-driven} methods. These can be based on different approaches, such as a principal component analysis, statistical pattern classifiers, neural networks, and support vector machines \cite{goetz2015evaluating}. Among FDI methods based on neural networks, Sorsa \emph{et al.}~\cite{sorsa1993dynamic} proposed the use of a multilayer perceptron, while Zhao \emph{et al.}~\cite{zhao2019neighborhood} introduced a new statistical feature extraction method named \emph{neighborhood-preserving neural network}. Other recent results include an FDI method that integrates an incremental one-class algorithm to identify anomaly conditions if an unknown fault occurs, and a dynamic shallow neural network for the classification of the fault state~\cite{arunthavanathan2020fault}, and another FDI method based on convolutional neural networks~\cite{chen2020data}.

In some cases, model-based techniques for FDI or FTC can include elements typical of data-driven approaches. For instance, a backstepping-based adaptive FTC method for Markov jump linear systems was introduced in \cite{yang2020neural}, where radial basis function neural
networks were used to model the unknown nonlinearities
in the system.

A recent trend in machine learning has been end-to-end learning, which condenses multiple stages of processing for a given task into a single deep neural network (see, e.g., \cite{amodei2016deep,tzirakis2017end}). A similar idea was applied to FTC in \cite{aznar2019obtaining}, in which a model was trained using reinforcement learning to directly handle the fault of ultrasound sensors of a mobile robot in a kinematic obstacle avoidance problem. Even if the approach developed in \cite{aznar2019obtaining} consisted of a sequence of deep neural networks with sensor measurements as input and robot action as output, the robot state was explicitly estimated as an intermediate variable.

In this work, we focus on end-to-end learning in FTC, and in particular on the case of abrupt sensor faults. The stages of FDI and control are replaced with a single recurrent neural network (RNN) with sensor measurements as input and control variables as output, in order to obtain a faster design process compared to classical methods. In contrast to \cite{aznar2019obtaining}, our deep FTC (DFTC) method has no explicit representation of the observed system states, and its training is based on supervised learning, rather than reinforcement learning. DFTC only requires (i) the availability of a (non-fault-tolerant) full state feedback control law, which is used as an ideal reference during the training phase, and (ii) the observability of the state vector using only the available non-faulty sensors, for all considered sensor faults. The model equations are not needed for designing our DFTC scheme, although it is assumed that a simulator for the system (e.g., a physics engine) is available in order to be able to train the control law; alternatively, one could directly use the actual system in real-life experiments, but this would cause issues related to safety and to the amount of time needed to obtain a sufficient amount of data for training. This type of approach, to the best of the authors' knowledge, has never been proposed in the literature and thus represents an innovative contribution. The main benefit of our approach as compared to classical model-based FTC schemes is the fact that a process model is not needed, and the definition of the DFTC parameters is entirely data-driven. Apart from making the FTC problem solvable in case a model is actually not available, our approach can also provide advantages when the classical FTC design (based on FDI, control reconfiguration block, and closed-loop controllers) can be difficult. This follows the general motivation for the use of data-driven control methods, which can have a \lq\lq major practical impact especially in those situations where identifying a process model can be difficult and time consuming, for instance, when data is affected by noise or in the presence of nonlinear dynamics'' \cite{de2019formulas}. On the other hand, the proposed approach presents some typical design challenges of methods based on neural networks, such as the need for a large amount of data and hyperparameter tuning. From a practical standpoint, the main contribution of the paper is the development of a deep learning pipeline for real-time closed-loop FTC. This pipeline includes dataset generation and augmentation, model training, and deployment of the model in a real-world setup.

The proposed FTC strategy is directly applied to a strongly nonlinear mechatronic system, consisting of a spherical inverted pendulum actuated by electric motors via reaction wheels. The control law to be imitated by the RNN was developed in \cite{baimukashev2020reaction} as an optimal control law.

The paper is organized as follows. Section \ref{sec:general_method} introduces the general approach, together with the required assumptions. Section \ref{sec:problem_form} describes the case study, while Sections \ref{sec:implem} and \ref{sec:DFTC_gen} provide information on the actual physical implementation (physical setup, hardware and software) and the generation of the DFTC, respectively. Simulation and experimental results are presented and discussed in Section \ref{sec:experim}, and finally conclusions are drawn in Section \ref{sec:concl}.

\section{End-to-end Deep Fault Tolerant Control}\label{sec:general_method}
The objective of this work is to design a closed-loop control law for a nonlinear dynamical system
\begin{subequations}\label{eq:system}
\begin{align}
    \dot{\boldsymbol{x}}&=\boldsymbol{f}(\boldsymbol{x},\boldsymbol{u})\label{eq:sys_seq}\\
    \boldsymbol{y}&=\boldsymbol{g}(\boldsymbol{x},\boldsymbol{u})\label{eq:sys_output}
\end{align}
\end{subequations}
where $\boldsymbol{x}\in\rr^{n_x}$ is the system state, $\boldsymbol{u}\in\rr^{n_u}$ is the control input, and $\boldsymbol{y}\in\rr^{n_y}$ is the system output, representing here the variables measured by $n_y$ sensors. Also, $\boldsymbol{f}(\cdot,\cdot):\rr^{n_x+n_u}\rightarrow\rr^{n_x}$ and $\boldsymbol{g}(\cdot,\cdot):\rr^{n_x+n_u}\rightarrow\rr^{n_y}$ are nonlinear functions defining the state and output equations, respectively. A control law is available for the system, as
\begin{equation}\label{eq:of}
    \boldsymbol{u}=\boldsymbol{k}(\boldsymbol{x}),
\end{equation}
where $\boldsymbol{k}(\cdot):\rr^{n_x}\rightarrow\rr^{n_u}$ is a nonlinear state-feedback control policy (referred to as \emph{baseline controller}) that aims at regulating the system state to the origin. 

\begin{assumption}\label{ass:nomodel}
The exact expressions of functions $\boldsymbol{f}(\boldsymbol{x},\boldsymbol{u})$, $\boldsymbol{g}(\boldsymbol{x},\boldsymbol{u})$, and $\boldsymbol{k}(\boldsymbol{x})$ are not necessarily available; however, we assume that it is possible to simulate both the open-loop system \eqref{eq:system} given an initial condition $\boldsymbol{x}_0$ and a control signal $\boldsymbol{u}$, and the closed-loop system \eqref{eq:system}-\eqref{eq:of} given an initial condition $\boldsymbol{x}_0$.\hfill$\square$
\end{assumption}

Assumption \ref{ass:nomodel} plays an important role, as, for many commercial mechatronic and robotic systems, software simulators are available for both the physical part of the system, and the controllers already implemented in the product; on the other hand, the equations for both are often not available, and their reverse engineering would require substantial time and effort.

We assume that $n_f$ different sensor faults can happen. When a fault affects a number $m^{(i)}$ of sensors, only the sensors associated with the remaining components of $\boldsymbol{y}$, namely $\boldsymbol{y}^{(i)}\in\rr^{n_y-m^{(i)}}$, provide useful information for control. The corresponding output equation, discarding faulty sensor measurements, will be
\begin{equation}\label{eq:output_fault}
    \boldsymbol{y}^{(i)}=\boldsymbol{g}^{(i)}(\boldsymbol{x},\boldsymbol{u}),\ i=1,\hdots,n_f,
\end{equation}
where $\boldsymbol{g}^{(i)}(\boldsymbol{x},\boldsymbol{u}):\rr^{n_x+n_u}\rightarrow\rr^{n_y-m^{(i)}}$ is simply obtained by discarding the elements of $\boldsymbol{g}(\boldsymbol{x},\boldsymbol{u})$ corresponding to faulty sensors.
\begin{assumption}\label{ass:obs} All of the $n_f$ systems given by the state equation \eqref{eq:sys_seq} and one of the $n_f$ output equations in \eqref{eq:output_fault} are fully observable. \hfill$\square$
\end{assumption}

As, according to Assumption \ref{ass:nomodel}, our approach does not require the availability of a model of the system, it is important to notice that Assumption \ref{ass:obs} can still be verified by employing numerical methods based on the evaluation of the observability Gramian matrix (see, e.g., \cite{empgram,vaidya2007observability}), as classical nonlinear observability analysis (based on Lie derivatives, as described for instance in \cite[Ch. 11]{marquez2003nonlinear}) would not be feasible. By following an approach similar to that employed in \cite{rakhim2019optimal}, for each of the $n_f$ considered cases, extensive simulations of the system dynamics are performed within a total time interval $T_o$, and the elements $W^{(i)}_{jk}$ of the observability Gramian $W^{(i)}\in\rr^{n_x\times n_x}$ (of row $j$ and column $k$) for the $i^{th}$ configuration are calculated as
\begin{equation*}
     W^{(i)}_{jk}\!=\! \frac{1}{4\epsilon^2} \int_0^{T_o}\! \left(\boldsymbol{y}^{(i)}_{+j}(t) - \boldsymbol{y}^{(i)}_{-j}(t)\right)^T\!\left(\boldsymbol{y}^{(i)}_{+k}(t) - \boldsymbol{y}^{(i)}_{-k}(t)\right)dt,
\end{equation*}
where $0<\epsilon\ll 1$, and $\boldsymbol{y}^{(i)}_{\pm j}$ is the sensor measurement generated by the simulation of the system dynamics for the $i^{th}$ configuration with initial condition $\bar{\boldsymbol{x}}_{\pm j} = \bar{\boldsymbol{x}} \pm \epsilon \boldsymbol{e}_j$, where $\boldsymbol{e}_j$ is the $j^{th}$ standard coordinate vector (i.e., the vector that contains all zeros, apart from the $j^{th}$ element, which is equal to $1$). Analogously, $\boldsymbol{y}^{(i)}_{\pm k}$ is the sensor measurement generated by the simulation of the system dynamics for the $i^{th}$ configuration with initial condition $\bar{\boldsymbol{x}}_{\pm k} = \bar{\boldsymbol{x}} \pm \epsilon \boldsymbol{e}_k$, where $\boldsymbol{e}_k$ is the $k^{th}$ standard coordinate vector. As observability measure for each configuration, we use
\begin{equation*}
    J^{(i)}\triangleq-log\left(det\left({W^{(i)}}^{-1}\right)\right)
\end{equation*}
with larger values of $J^{(i)}$ indicating \lq\lq better'' observability.

In the presence of sensor faults, the method described in this work aims at synthesizing a DFTC law
\begin{equation}\label{eq:kft}
    \boldsymbol{u}=\boldsymbol{k}_{ft}(Y_T)
\end{equation}
where $Y_T$ represents the available (possibly faulty) measurements of $\boldsymbol{y}$ in the time interval $T$ immediately preceding the current time instant. The function $\boldsymbol{k}_{ft}(\cdot)$ is implemented using a RNN, whose realization is obtained by following a procedure consisting of two phases. In the first phase, the closed-loop system \eqref{eq:system}-\eqref{eq:of} is initialized at several initial conditions $\boldsymbol{x}_0$, randomly determined in a suitable region of the state space. The time evolution of $\boldsymbol{y}$ is thus recorded from the simulated closed-loop system \eqref{eq:system}-\eqref{eq:of}. In the second phase, the RNN is trained aiming at minimizing the difference between the control input for the closed-loop system \eqref{eq:system}-\eqref{eq:of} in the absence of sensor faults and the generated control input for the closed-loop system \eqref{eq:system}, \eqref{eq:kft} in the presence of sensor faults, which are randomly introduced in the system. The DFTC law thus learns to reproduce the control variable of the fault-free baseline controller even in the case of faulty sensors.

%In the second phase, the RNN is trained aiming at minimizing the difference between the recorded state evolution for the closed-loop system \eqref{eq:system}-\eqref{eq:of} in the absence of sensor faults, and the state evolution for the closed-loop system \eqref{eq:system}, \eqref{eq:kft} in the presence of sensor faults, which are randomly introduced in the system.

Let us refer to the nominal evolution of the state for system \eqref{eq:system}-\eqref{eq:of} as $\boldsymbol{x}(t)$. %, and to the evolution of the state of \eqref{eq:system}, \eqref{eq:kft} from the same initial condition, in the possible presence of sensor faults, as $\tilde{x}(t)$. %Also, the output measurements (again, including possible sensor faults) corresponding to $\tilde{x}(t)$ is referred to as $Y_T(t)$. 
Given $n$ time instants $t_i$ for which the values of $\boldsymbol{k}(\boldsymbol{x}(t_i))$ have been recorded during the first phase, the RNN will be trained in the second phase so as to minimize the loss function
\begin{equation}\label{eq:cost_general}
    P\triangleq \frac{1}{n}\sum_{i=1}^n d\left(\boldsymbol{k}(\boldsymbol{x}(t_i)),\boldsymbol{k}_{ft}(Y_T(t_i))\right),
\end{equation}
where, e.g., a squared loss $d\left(\boldsymbol{a},\boldsymbol{b}\right)\triangleq\left(\boldsymbol{a}-\boldsymbol{b}\right)^T\left(\boldsymbol{a}-\boldsymbol{b}\right)$ can be employed, but one can choose other options such as absolute loss or Huber loss functions.

The DFTC pipeline is presented in Algorithm \ref{alg:dl_pipeline}. First, in the pre-processing phase, a dataset is generated using the baseline controller for different initial conditions of the system states, without sensor faults. Then, the dataset is augmented by adding synthetic sensor faults. During the training phase, the network is trained until convergence by updating its weights at each iteration using back-propagation. Afterwards, in the inference phase, the trained network is deployed in a real-world setup to stabilize the system via closed-loop control. 

% algorithm
\begin{algorithm}[t!] 
\SetAlgoLined
\SetKwInOut{Input}{input}\SetKwInOut{Output}{output}
\textbf{Pre-processing phase} \\
Generate data using the baseline controller\\
Augment the data by adding sensor faults \\
Divide the data into train, test, and validation sets\\
\textbf{Training phase} \\
Define the network architecture \\
Set the network hyperparameters \\

\For(\tcp*[h]{number of epochs} ){$i=0$ \KwTo $M$ }{  
    \For(\tcp*[h]{number of batches} ){$j=0$ \KwTo $S$}{ 
        Randomly sample batch of inputs \\
        Perform feed-forward pass \\ 
        Calculate the predicted output \\ 
        Compute the loss \\   
        Perform back-propagation \\ 
        Update weights of the network \\  
    }
}

\textbf{Inference phase} \\
Load the trained network \\
\For{$t_i=t_0$ \KwTo $t_n$}{
    Read the sensor values \\
    %Introduce the fault to sensor measurements \\
    Perform feed-forward pass\\
    Calculate the predicted output \\ 
    Apply control inputs \\
}

\caption{DFTC Pipeline}
\label{alg:dl_pipeline}
\end{algorithm}

Contrary to classical FTC schemes, where the state estimates generated by the FDI system are passed to the FTC scheme, the methodology defined in this paper directly determines the control action $\boldsymbol{u}$ as in \eqref{eq:kft}, without explicitly trying to reconstruct state estimates. This is why the proposed FTC scheme has an end-to-end nature.

For the reader's convenience, Table \ref{tab:notations} presents the alphabetically sorted list of symbols of the general approach described in Section \ref{sec:general_method} and used throughout the paper.

\begin{table}[ht]
\centering % centering table
\begin{tabular}{ll} % 4 columns
\toprule
\text{Symbol}                &  Description \\
\midrule
$\boldsymbol{f}$       & vector of state equations   \\
$\boldsymbol{g}$       & vector of output equations   \\
%$J$         & cost function of a trajectory \\
$J^{(i)}$       & observability measure of $i^{th}$ configuration   \\
$\boldsymbol{k}$       & baseline control law   \\
$\boldsymbol{k}_{ft}$ & DFTC law\\
$n$         & batch size for training                 \\
$n_f$         & number of possible sensor faults            \\
$P$         & training loss function        \\
%$Q$         & weights of the states         \\
%$R$         & weights of the control inputs \\
$\boldsymbol{x}$         & system state                  \\
$\boldsymbol{y}$         & system output (sensor measurements)           \\
$\boldsymbol{y}^{(i)}$         & $i^{th}$ system output vector excluding faulty sensors           \\
$Y_T$         & available measurements in the time interval $T$           \\
$\boldsymbol{u}$         & control input                 \\
$W^{(i)}$         & observability Gramian of the $i^{th}$ configuration\\
%$\lambda$   & regularization parameter      \\
%$\rho$      & normalized cost               \\

\bottomrule
\end{tabular}
\caption{General notation}
\label{tab:notations}
\end{table}

\section {Case Study: Problem Formulation}\label{sec:problem_form}
The case study considered in this paper focuses on the FTC of an inverted pendulum actuated by two reaction wheels. In this section, we provide a model, in form \eqref{eq:system}, of the physical system (Section \ref{sec:sys_mod}), the definition of the state feedback control law as in \eqref{eq:of} (Section \ref{sec:clcontr}), an analysis of the admissible sensor faults that define equation \eqref{eq:output_fault}, the results of the observability analysis of the system (both in Section \ref{sec:sensfaults}), and a general description of the end-to-end FTC architecture that will generate $\boldsymbol{k}_{ft}(Y_T)$ (Section \ref{sec:DL}).

\subsection{System Modeling}\label{sec:sys_mod}
\begin{figure}[tbp]
    \centering
    \begin{subfigure}[b]{.49\columnwidth}
    \centering
    \includegraphics[width=3.4cm]{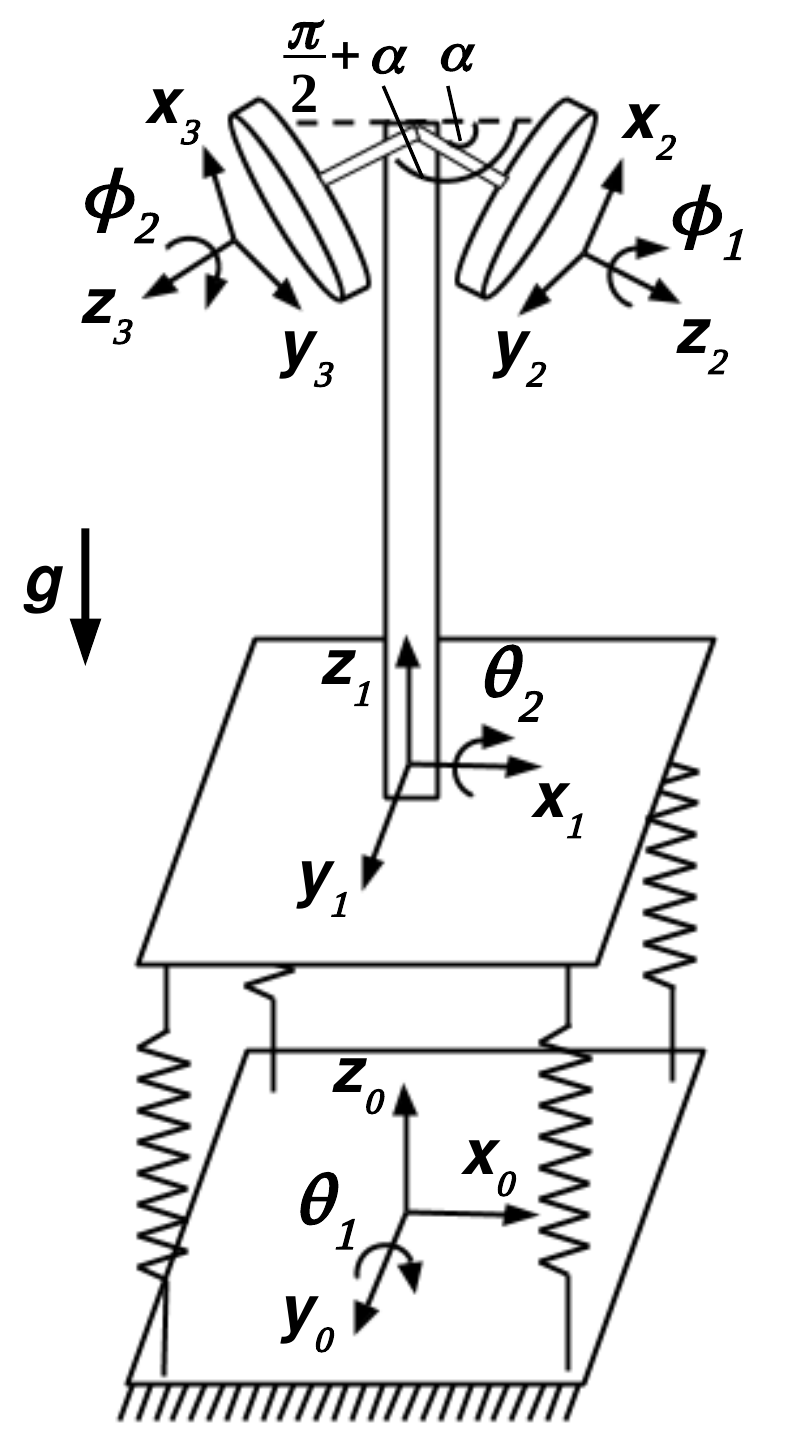}
    \caption{Block diagram.}
   % \label{fig:block_diagram}
    \end{subfigure}%
    \begin{subfigure}[b]{.49\columnwidth}
    \centering
    \includegraphics[width=3.9cm]{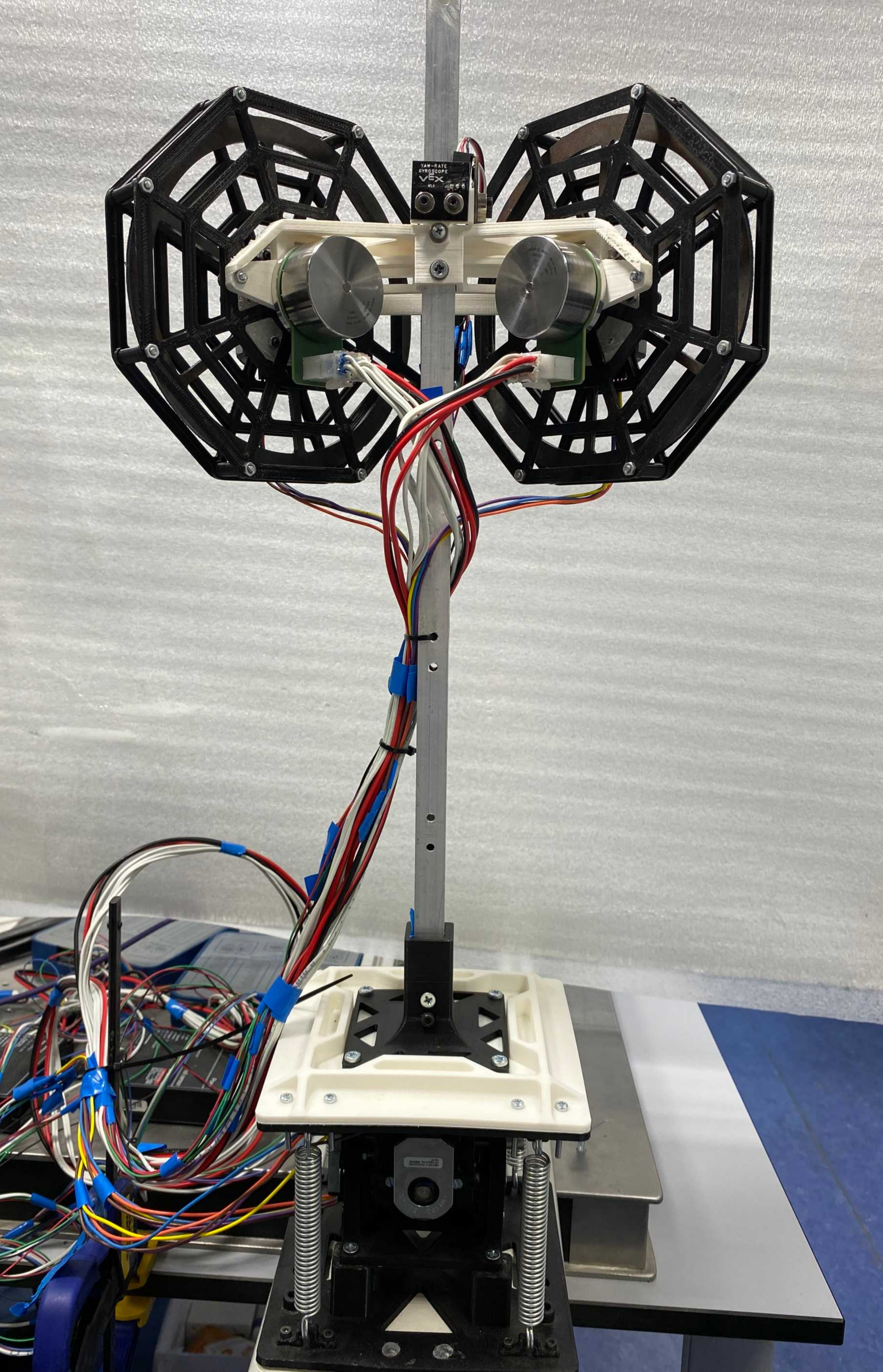}
    \caption{Reaction wheel setup.}
   % \label{fig:setup_overall}
    \end{subfigure}
    \caption{Schematics and real-world setup of the inverted pendulum with dual-axis reaction wheels \cite{baimukashev2020reaction}.}
    \label{fig:block_and_setup_overall}
\end{figure}

 The schematics of the inverted pendulum and the corresponding real-world setup are shown in Fig. \ref{fig:block_and_setup_overall}. The system configuration, defined as $\boldsymbol{q}=\begin{bmatrix}\theta_1\ \theta_2\ \phi_1\ \phi_2\end{bmatrix}^T\in\rr^4$ is used to describe the equations of motion. In this representation, $\theta_1$ and $\theta_2$ describe the angular rotation of the pendulum, while $\phi_1$ and $\phi_2$ are the angles of rotation, respectively, of each of the two reaction wheels.

The detailed derivation of the equations of motion using the Lagrangian formulation is presented in \cite{baimukashev2020reaction}, and leads to
\begin{equation}
    M(\boldsymbol{q})\ddot{\boldsymbol{q}} + C(\boldsymbol{q},\dot{\boldsymbol{q}}) + G({\boldsymbol{q}}) = \boldsymbol{\tau},
  \label{eq:motion_equation}
\end{equation}
where $M(\boldsymbol{q})\in\rr^{2\times 4}$ is the rigid body inertia matrix, $C(\boldsymbol{q},\dot{\boldsymbol{q}})\in\rr^{2}$ accounts for Coriolis and centrifugal forces, $G({\boldsymbol{q}})\in\rr^{2}$ represents the forces due to gravity and the elastic forces acting on the system, and $\boldsymbol{\tau}\in\rr^2$ is the vector of external torques generated by the motors that actuate each of the reaction wheels. In turn $\boldsymbol{\tau} = k_T[i_1\ i_2]^T$, where $i_1$ and $i_2$ denote the currents of the motors that actuate the two reaction wheels, and $k_T$ is the torque constant of both motors. The reader is referred to \cite[Sec. II]{baimukashev2020reaction} for a precise expression of these functions, and to Table I in \cite{baimukashev2020reaction} for the list of system parameters with precise physical values.

The state of the system is thus introduced as $\boldsymbol{x}=\begin{bmatrix}\theta_1 & \theta_2 & \dot{\theta}_1 & \dot{\theta}_2 & \dot{\phi}_1 & \dot{\phi}_2\end{bmatrix}^T\in\rr^6$, while the control input is $\boldsymbol{u}=\begin{bmatrix}i_1 & i_2\end{bmatrix}^T\in\rr^2$. The state-space representation in form \eqref{eq:sys_seq} is
\begin{subequations}\label{eq:seven}
\begin{equation}
    \dot{\boldsymbol{x}} = \boldsymbol{f}(\boldsymbol{x},\boldsymbol{u}) = \begin{bmatrix}\dot{\theta}_1\\ \dot{\theta}_2\\ \ddot{\boldsymbol{q}}\end{bmatrix}\in\rr^6,
    \label{eq:state_equation}
\end{equation}
where $\ddot{\boldsymbol{q}}=M^{-1}(\boldsymbol{\tau} - C - G)\in\rr^4$ is obtained from \eqref{eq:motion_equation}.

Each system state is measured by a separate sensor: in particular, four optical encoders are used to measure $\theta_1$, $\theta_2$, $\dot{\phi}_1$ and $\dot{\phi}_2$, while two gyro sensors measure $\dot{\theta}_1$ and $\dot{\theta}_2$. The output equation \eqref{eq:sys_output} is thus defined by $\boldsymbol{g}(\boldsymbol{x},\boldsymbol{u})=\boldsymbol{x}$, i.e.,
\begin{equation}
    \boldsymbol{y} = \boldsymbol{x}.
    \label{eq:output_equation}
\end{equation}
\end{subequations}

\subsection{State-Feedback Control Law}\label{sec:clcontr}
The baseline controller $\boldsymbol{k}(\boldsymbol{x})$ in \eqref{eq:of} is obtained by solving a numerical optimal control problem (OCP) within a fixed time horizon, which approximates an infinite-horizon optimal control law. As the focus of this work is not on the formulation of the baseline controller, the reader is referred to \cite[Sec. III.A]{baimukashev2020reaction} for the details of its design process. It is important to highlight that the baseline controller has no fault tolerance properties, and was designed for full state feedback. Finally, notice that, given the availability of analytical expressions of functions $\boldsymbol{f}(\boldsymbol{x},\boldsymbol{u})$ and $\boldsymbol{g}(\boldsymbol{x},\boldsymbol{u})$ in \eqref{eq:seven} as a particular realization of \eqref{eq:system}, and the possibility of obtaining $\boldsymbol{k}(\boldsymbol{x})$ numerically for any state value, it is possible to simulate both system \eqref{eq:system} for a given initial condition and control signal, and system \eqref{eq:system}-\eqref{eq:of} for a given initial condition: this implies the satisfaction of Assumption~\ref{ass:nomodel}.

\subsection{Sensor Faults and System Observability}\label{sec:sensfaults}
Each of the previously mentioned sensors can be subject to faults during system operation. In particular, we assume that three types of abrupt sensor fault might happen: (i) the sensor provides a constant value, which corresponds to the state measurement at a given time instant $t$, but remains unchanged while the state value varies after $t$; (ii) the sensor suddenly outputs a zero measurement; (iii) the sensor suddenly outputs a constant value, corresponding to its maximum range. The reader is referred to \cite{li2020recent,fan2012sensor} for more details on abrupt sensor failures. In our system, the most critical situation would happen in case of a fault in one of the encoders measuring $\dot{\phi}_1$ or $\dot{\phi}_2$, as this would lead to very high rotational speeds, possibly with destructive consequences.

Under the reasonable assumption that a fault can only happen for one sensor at a time, we obtain a realization of \eqref{eq:output_fault} with $n_f=n_y=6$, and $\boldsymbol{y}^{(i)}\in\rr^5$ defined as vector $\boldsymbol{y}\in\rr^6$, without the element of position $i$. In order to be able to design our DFTC law, we need to guarantee that each of these $n_f=6$ configurations results in an observable system, as required by Assumption \ref{ass:obs}.

Given the complexity of the considered system, which prevented us from using classical methods for observability analysis despite the availability of an analytical model, the numerical method based on the observability Gramian described in Section \ref{sec:general_method} was employed. After running extensive simulations, a value of $J^{(i)}$ was obtained for each of the $n_f=6$ cases, which measured the observability when the $i^{th}$ sensor was faulty. Table \ref{tab:my_label} summarizes the results of this test. The value of $J^{(i)}$ for each configuration $\boldsymbol{y}^{(i)}$ is reported, together with the value of the same quantity when the full output vector $\boldsymbol{y}$ is available. We notice that the value of $J^{(i)}$  for all considered configurations is only slightly below the ideal case: in fact, the latter has an observability measure of 31.2, with the same measure never decreasing below 29.8 for all other configurations. For comparison, the best possible configuration using only two sensors has an observability measure of $12.5$. We can thus conclude that Assumption \ref{ass:obs} is satisfied for the considered system.

\begin{table}[ht]\label{obs_table}
    \centering
    \begin{tabular}{|c|c|c|}
        \hline
         \textbf{Output}& \textbf{Available sensors} & $\boldsymbol{J^{(i)}}$\\
         \hline
          $\boldsymbol{y}^{(1)}$ &[2,3,4,5,6] & 29.9\\
         \hline
         $\boldsymbol{y}^{(2)}$ &[1,3,4,5,6] & 30.1\\
         \hline
          $\boldsymbol{y}^{(3)}$ &[1,2,4,5,6] & 29.8 \\
         \hline
          $\boldsymbol{y}^{(4)}$ & [1,2,3,5,6] & 30.0\\
         \hline
          $\boldsymbol{y}^{(5)}$ & [1,2,3,4,6] & 30.5\\
         \hline
          $\boldsymbol{y}^{(6)}$ & [1,2,3,4,5] & 30.1\\
         \hline
         $\boldsymbol{y}$ & [1,2,3,4,5,6] & 31.2 \\
         \hline
    \end{tabular}
    \caption{Observability measure for different sensor configurations.}
    \label{tab:my_label}
\end{table}

\subsection{DFTC Architecture}\label{sec:DL}

\begin{figure}[b!]{}
    \centering
   \includegraphics[width=0.95\columnwidth]{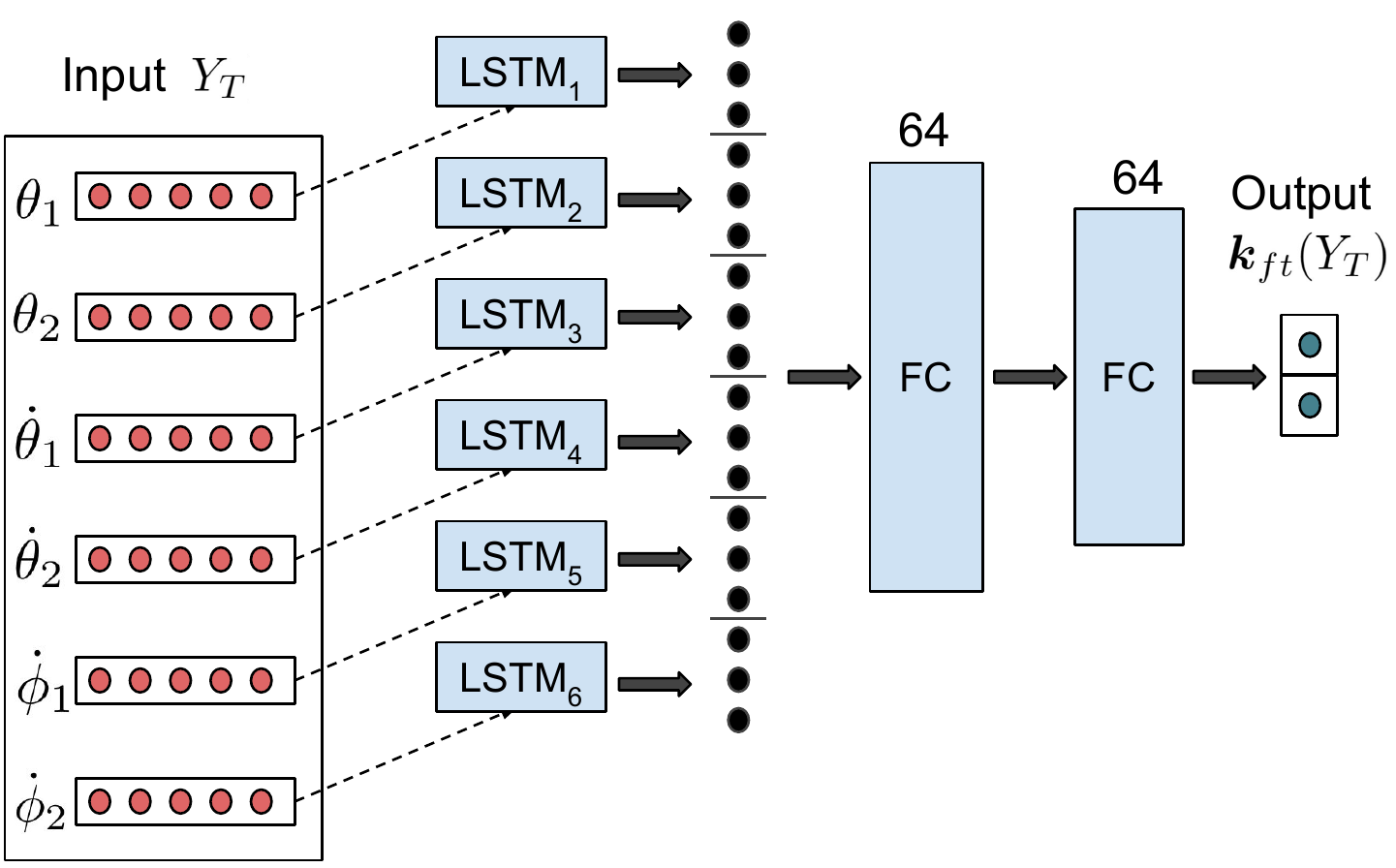}
    \caption{Architecture of the RNN that implements DFTC.}
    \label{fig:model}
\end{figure}

To generate the control law $\boldsymbol{u}=\boldsymbol{k}_{ft}(Y_T)$ introduced in \eqref{eq:kft}, the deep neural network that implements the DFTC was trained via supervised learning to imitate the OCP trajectories. The deep neural network, represented in Fig. \ref{fig:model}, consists of six shallow long short-term memory (LSTM) blocks followed by two fully-connected (FC) layers. We chose to utilize RNNs to process sensor signals because of their continuous nature. The $i^{th}$ LSTM block takes as input the measured values of the $i^{th}$ output signal for the last $m$ time steps. Then, the outputs of all LSTM blocks are concatenated, and the extracted features are passed to a FC network with two layers. The output of the network consists of the two components of the control input vector $\boldsymbol{u}$. The cost function for model training $P$ is defined, with reference to \eqref{eq:cost_general}, as the following quadratic loss function:
% with one layer and 32 hidden cells
% with 64 nodes each
%\begin{equation}%\label{eq:loss}
%P\!=\!\frac{1}{2n}\sum_{i=1}^{n}{\left(k(x(t_i))\!-\!k_{ft}(Y_T(t_i))\right)^T\!\left(k(x(t_i))\!-\!k_{ft}(Y_T(t_i))\right)}
%\end{equation}
\begin{equation*}
P=\frac{1}{2n}\sum_{i=1}^{n}{\boldsymbol{v}_i^T\boldsymbol{v}_i},
\end{equation*}
where $\boldsymbol{v}_i\triangleq \boldsymbol{k}(\boldsymbol{x}(t_i))\!-\!\boldsymbol{k}_{ft}(Y_T(t_i))$.

It is worth remarking that a single multi-input multi-output LSTM block was also tested as a possible solution instead of one LSTM block for each system output. This type of architecture learned to rely mainly on link velocity measurements: as a consequence, the closed-loop system showed oscillations in case of link velocity faults. We thought that this can be a problem in several different applications, and thus decided to rely on the described architecture with $n_y$ LSTM blocks.

\section {Case Study: Implementation}\label{sec:implem}

\subsection{{Physical Setup}}
The inverted pendulum with dual-axis reaction wheels has a universal joint in the center of its base, which enables the rotation of the pendulum around the pivot point (see Fig.~\ref{fig:block_and_setup_overall}). Four extension springs attached to the corners of the base keep the pendulum in the vertical position. The pendulum consists of an aluminum bar and two reaction wheels attached to it and placed orthogonal to each other. The lightweight structure that connects the motors and reaction wheels to the link was printed using Ultimaker S5 with PLA material. The reaction wheels were fabricated from a steel plate with a thickness of 3 mm using a laser cutter (Bodor P3015). Each flywheel has a 160 mm diameter and weighs 160 grams. 

\subsection{Hardware}

Two flat brushless DC motors (EC45, Maxon Motors) with a 134 mNm nomimal torque were selected for actuating the reaction wheels. The motors were controlled by two servoamplifiers (DEC 50/5, Maxon Motors) in current mode. The angular rotation of the pendulum with respect to the vertical axis and reaction wheel velocities were measured using capacitive incremental encoders (AMT 102, CUI Devices). Two analog gyroscope sensors (LY3100ALH, STMicroelectronics) were used to measure the link velocities in two axes. 
%We measured the velocities using analog gyroscopess instead of taking the numerical derivatives of the encoders since the velocities obtained from the encoders the nume
A workstation (Intel i7-4790 processor, 32 GB memory, CentOS 7 operating system) with two data acquisition cards (PCI-6259, National Instruments) was used to read the sensor signals and to send the control commands to the servo amplifiers. The real-time calculation of the DTFC was performed using a graphics processing unit (GeForce GTX 1080, NVIDIA).

\subsection{Software}
A number of trajectories of system \eqref{eq:seven} with the baseline controller $\boldsymbol{k}(\boldsymbol{x})$ described in Section \ref{sec:clcontr} were simulated using Matlab 9.7. Afterwards, the data preprocessing and model training pipeline was implemented in Python 3.5. The deep neural network architecture was built using the PyTorch library. The trained model was then converted to TorchScript that can be run on different environments, regardless of Python dependencies. This enables the model deployment in the C++ environment for real-time operation. In particular, the \textit{LibTorch} library was used to integrate the model into the C++ script. In the experimental setup, closed-loop control was implemented using the C++11 language. This program acquired the sensor measurements and sent the control signals to the servo amplifiers through the \textit{Ni-DAQmx} driver. The system was run on the CentOS 7 operating system. 
\section{Case Study: DFTC Generation}\label{sec:DFTC_gen}

%In our model, we use knowledgebased fault tolerance technique. Artificial neural network is used to control the behavior of the system in end to end fashion. In order to implement sensor fault tolerant, the following techniques is proposed. OCP data were generated for ideal case, without any fault in the sensors. Later more data are generated by modifying generated detest for ideal case. From states of the system, sensor data are generated as input data with abrupt sensor fault.

\subsection{Dataset Generation}\label{sec:datagen}

DFTC is a data-driven technique which uses supervised learning, thus requiring a dataset to be created. To this end, the same OCP mentioned in Section \ref{sec:clcontr} was solved from 4500 initial conditions $\boldsymbol{x}_0\in\rr^6$, uniformly sampled from a set defined by $\theta_1\in[-\pi/4,\pi/4]$ rad, $\theta_2\in[-\pi/4,\pi/4]$ rad, $\dot{\theta}_1\in[-6.8,6.8]$ rad/s, $\dot{\theta}_2\in[-6.8,6.8]$ rad/s, $\dot{\phi}_1\in[-300,300]$ rad/s, $\dot{\phi}_2\in[-300,300]$ rad/s. %The OCP was solved using We used OSQP\cite{banjac2017embedded} with multiple shooting approach. 
These values were chosen such that the resulting state evolutions of the system would include all representative motions in the workspace. The cost function minimized by the optimal control sequence is defined as follows:
\begin{equation}\label{eq:OCP_cost1}
	J_{\rm OCP}= \int_0^{4} \left(\boldsymbol{x}(t)^TQ\boldsymbol{x}(t)+\boldsymbol{u}(t)^TR\boldsymbol{u}(t)\right) dt,
\end{equation}
where $Q=\mathrm{diag}\left\{5\cdot 10^4,5\cdot 10^4,5\cdot 10^2, 10^2,10^{-2},10^{-2}\right\}$ and $R=\mathrm{diag}\left\{10^{-5},10^{-5}\right\}$, as in \cite{baimukashev2020reaction}.

Each simulation had a duration of $4$ s, chosen via trial and error by observing the settling time of the closed-loop system. A sampling interval of $10$ ms was used, resulting in $400$ time steps for each simulation, and a total number of $4500\cdot 400=1.8\cdot 10^6$ data points $(\boldsymbol{x},\boldsymbol{u}=\boldsymbol{k}(\boldsymbol{x}))$. The computation of all trajectories (Line 2 in Algorithm \ref{alg:dl_pipeline}) took around 70 hours on the workstation described in Section IV.C.

\subsection{Dataset Augmentation with Sensor Fault Scenarios}
The trajectories described in Section \ref{sec:datagen} refer to the ideal case when the values of the states are correctly measured and fed back to the baseline control law.

In order to train the RNN that implements our DFTC law, the dataset was augmented by mimicking abrupt sensor faults. Sensory faults in each trajectory were generated at a random time step in the interval $[0.3 s, 2 s]$. The sensor output for position sensor faults was set to a constant value in the interval $\pm \pi$, equal to the last measured value, or to zero. Instead, for velocity measurements, the same quantity was set to zero. For each trajectory in the dataset, we generated two abrupt sensor failures, resulting in 9000 augmented trajectories with synthetic faults. These data samples were added to the original 4500 trajectories, resulting in a dataset with 13500 trajectories. This operation, corresponding to Line 3 in Algorithm \ref{alg:dl_pipeline}, took 37 seconds.

\subsection{DFTC Model and Training}

%Specifically, in the OCP trajectory, the system is simulated for 4 seconds from the different initial conditions. Also, on average the system stablizes after around 2s from the beginning of simulation. Therefore, we crop the first 2.5 seconds of each trajectory and use it as training data. OCP solves the problem of generating the data-set, where many points are around the origin $x(t) \approx \textbf{0}$. 

The dataset was divided into training, validation and testing sets in the following proportions: training (80\%), validation (10\%) and testing (10\%). Dividing the data into three sets with randomly shuffled samples (Line 4 in Algorithm \ref{alg:dl_pipeline}) took 217 seconds. Each sample consisted of six state measurements for the last 10 time steps (i.e., $T = 100$ ms). During training, L2 regularization was used to prevent overfitting. The L2 regularized cost function became
\begin{equation*}
P_{L2} = P + \frac{1}{2n}\lambda\sum_{i=1}^{r}{V_{i}^2},
\end{equation*}
where $n$ is the batch size already introduced to define $P$ in \eqref{eq:cost_general}, $V_{i}$ are the weight parameters of the model, $r$ is the total number of parameters and $\lambda$ is the regularization parameter.

\begin{figure}[b!]{}
    \centering
   \includegraphics[width=1.0\columnwidth]{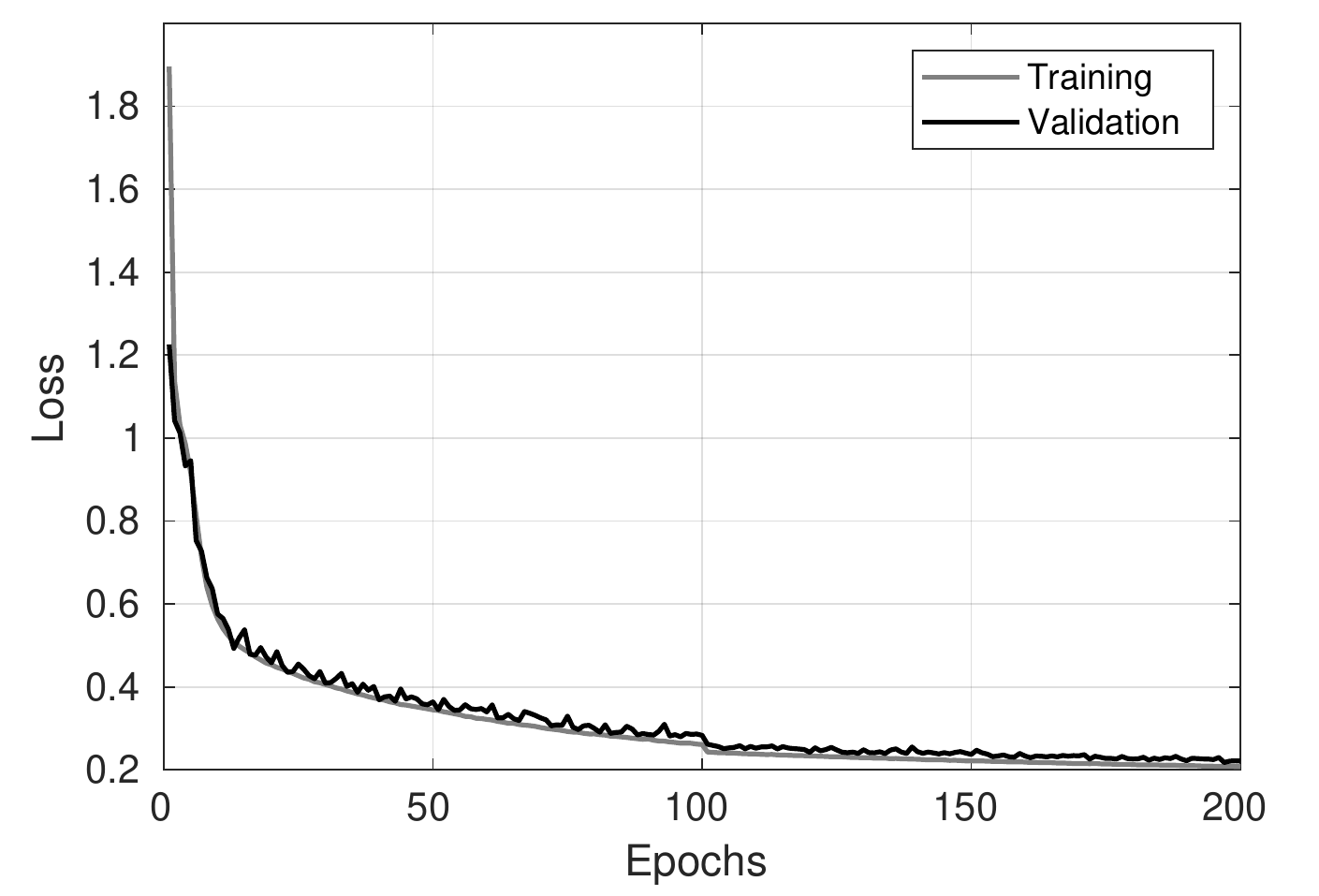}
    \caption{Learning curve, which shows the evolution of the training and validation loss with respect to training epochs.}
    \label{fig:train_curve}
\end{figure} 

Each LSTM block of the DFTC model had one layer and 32 hidden layer units, while both FC layers had 64 nodes and a ReLU activation function was used after them. The RMSprop optimizer with an initial learning rate of 0.001 was used. The learning rate was decreased to $5\cdot 10^{-4}$ after 100 epochs. The model was trained for 200 epochs on an Nvidia DGX-2 server, which took around 6 hours (with reference to Lines 5-17 in Algorithm \ref{alg:dl_pipeline}). The batch size $n$ was set to 1024, and the regularization parameter $\lambda$ was assigned the value of $10^{-3}$. The convergence of the training and validation loss without overfitting can be inferred from Fig.~\ref{fig:train_curve}.

\section{Experimental Results and Discussion}\label{sec:experim}

\subsection{Simulation Results}
\begin{figure*}
        \includegraphics[width=\linewidth]{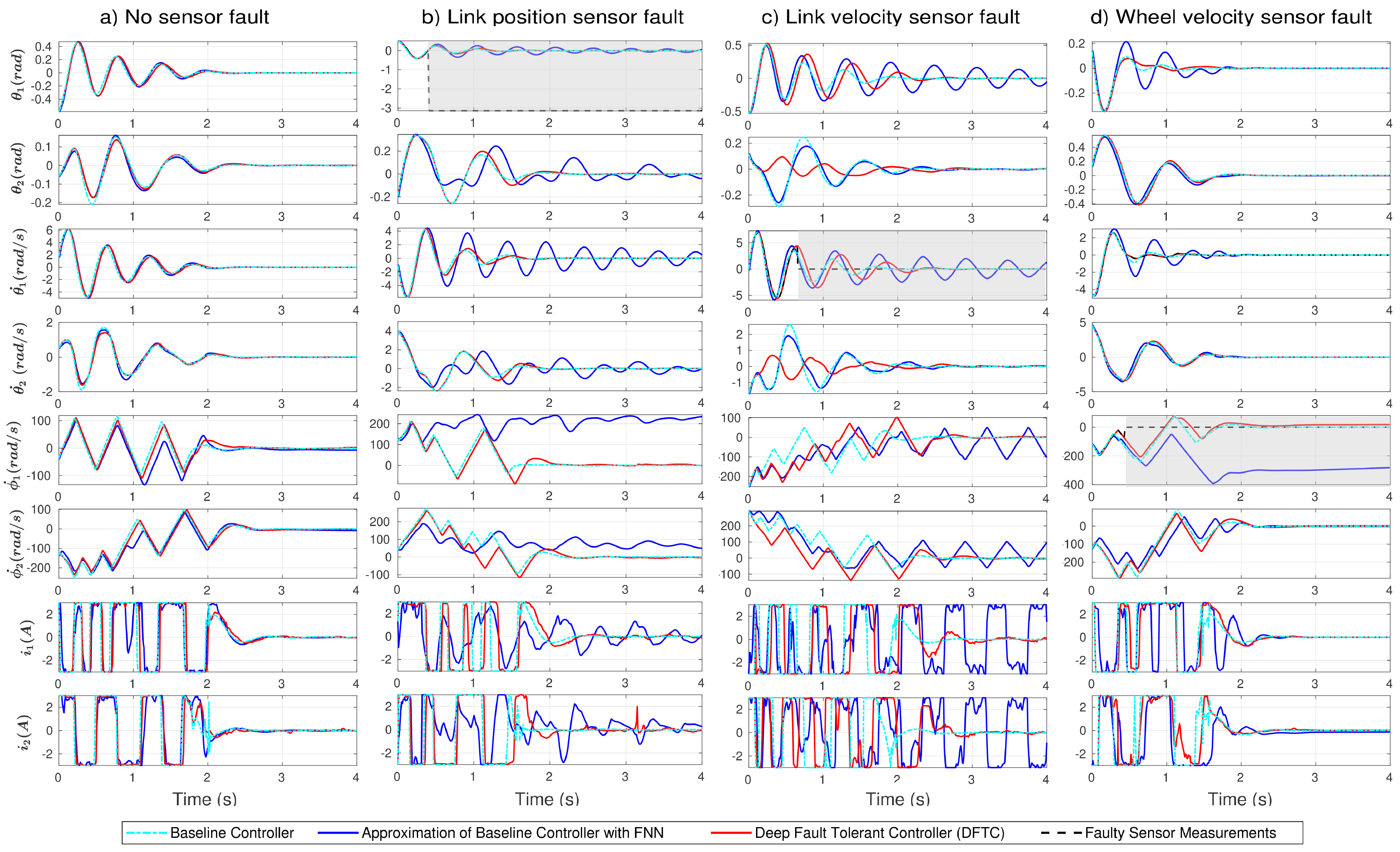}
        \caption{Simulation of time evolution of states and inputs for the baseline controller (always without faults), FNN and DFTC with different sensor fault scenarios.}
        \label{fig:sim_results}
\end{figure*}

The performance of the DFTC was compared against the approximation of the baseline controller via the feedforward neural network (FNN) described in \cite{baimukashev2020reaction}. This controller did not present fault-tolerance properties, but (contrary to the actual OCP-based baseline controller) could be implemented in real time. Specifically, two simulation scenarios were considered. In the first set of simulations, no sensor fault was present and all states of the system were available for closed-loop control. On the contrary, in the second set of simulations, random abrupt sensor failures were introduced in the time interval between $0.3$ s and $2$ s of the simulation. Five hundred pairs of simulations were run for a duration of $4$ s with a sampling time of $10$ ms from 500 different initial conditions. Then, for each DFTC and FNN simulation, the costs $J_{\rm DFTC}$ and $J_{\rm FNN}$ were calculated using the same formula defined for $J_{\rm OCP}$ in \eqref{eq:OCP_cost1}, by evaluating them a posteriori on simulation data. The normalized costs $\rho_{\rm DFTC}=J_{\rm DFTC}/J_{\rm OCP}$ and $\rho_{\rm FNN}=J_{\rm FNN}/J_{\rm OCP}$ were used to evaluate the loss of performance of the two deep-learning-based controllers as compared to the baseline control law.

The results of these simulations are summarized in Table~\ref{tab:rnn_cost}, while Fig.~\ref{fig:sim_results} presents the results of four sets of simulations with and without sensor faults. The failure of link position and velocity, and wheel velocity sensors are considered. For each of the four pairs of simulations, the results of the baseline controller, FNN and DFTC for the same initial conditions are compared.

\begin{table}[bp]
\centering
\begin{tabular}{|c|c|c|c|c|c|c|} 
\hline
\multirow{2}{*}{} & \multicolumn{2}{l|}{\begin{tabular}[c]{@{}l@{}} \textbf{Normalized cost}\\\textbf{without fault}\end{tabular}} & \multicolumn{2}{l|}{\begin{tabular}[c]{@{}l@{}} \textbf{Normalized cost}\\\textbf{with fault}\end{tabular}} & \multicolumn{2}{l|}{ \textbf{Execution Time}}  \\ 
\cline{2-7}
                & mean & std. dev & mean & std. dev & mean & worst \\
\hline
\textbf{DFTC} & 1.08 & 0.07 & 1.10 & 0.09 & 1.4 ms & 2.6 ms\\
\hline
\textbf{FNN} & 1.07 & 0.09 & 2.25 & 5.09 & 0.2 ms & 0.4 ms\\ 

\hline
\end{tabular}

\caption{Performance comparison of DFTC and FNN controller with and without sensor faults. The costs of DFTC and FNN were normalized by dividing the cost of the baseline controller without fault.}
\label{tab:rnn_cost}
\end{table}

\begin{figure*}
    \centering
    \includegraphics[width=\linewidth]{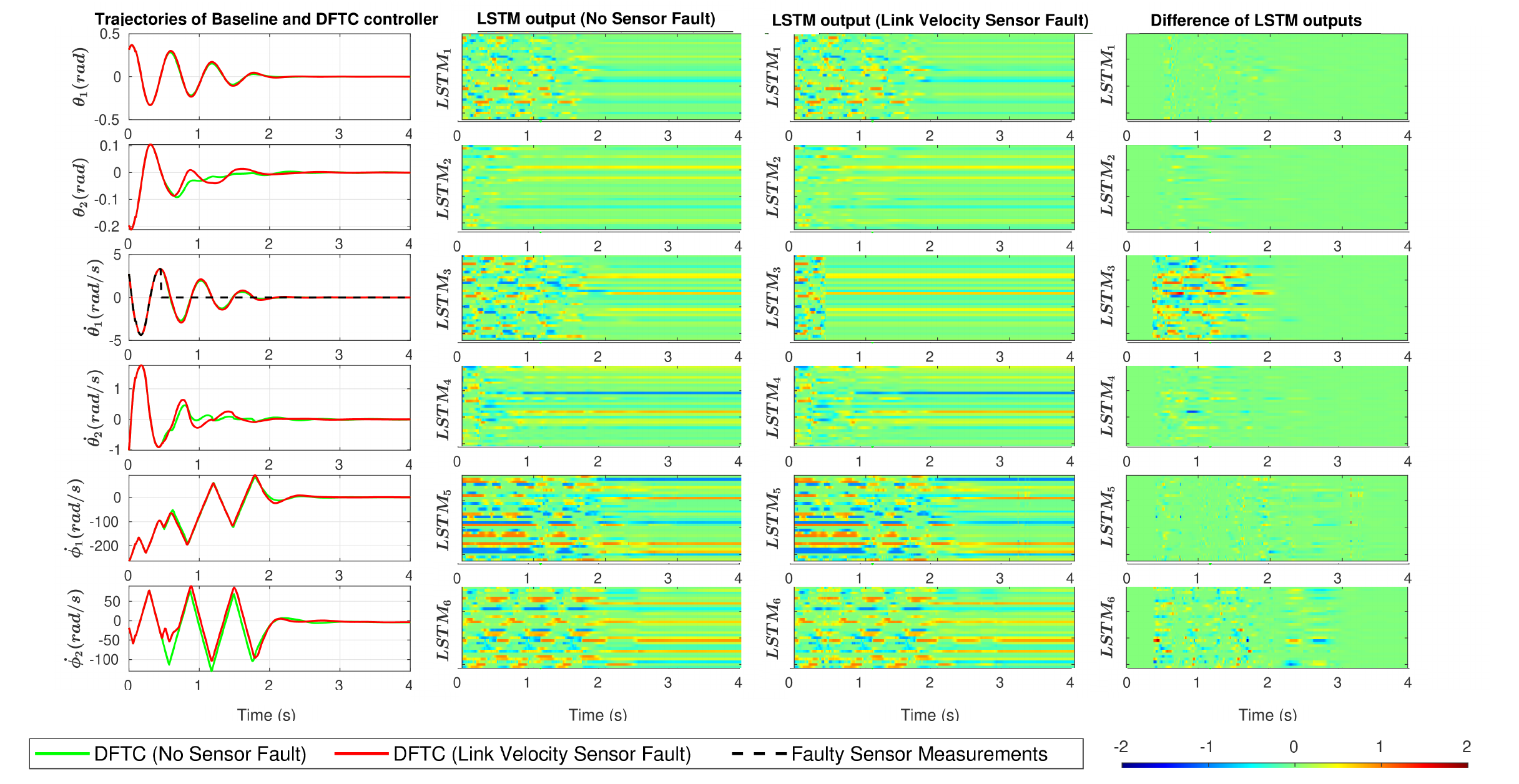}
    \caption{Output of LSTM block for respective input state with and without sensor fault measurements.}
    \label{fig:lstm_feat}
\end{figure*}

We also conducted two simulations to compare the state evolution of DFTC with a link velocity sensor fault and without a fault from the same initial condition. The states and the corresponding LSTM block outputs for both cases and their difference are shown in Fig.~\ref{fig:lstm_feat}. DFTC showed good control performance in both cases. Notice that the LSTM block output for the link velocity input remained unchanged after the abrupt sensor fault because a constant (faulty) zero velocity value was fed to it until the end of the simulation.

\subsection{Experimental Results}

\begin{figure*}
        \includegraphics[width=\linewidth]{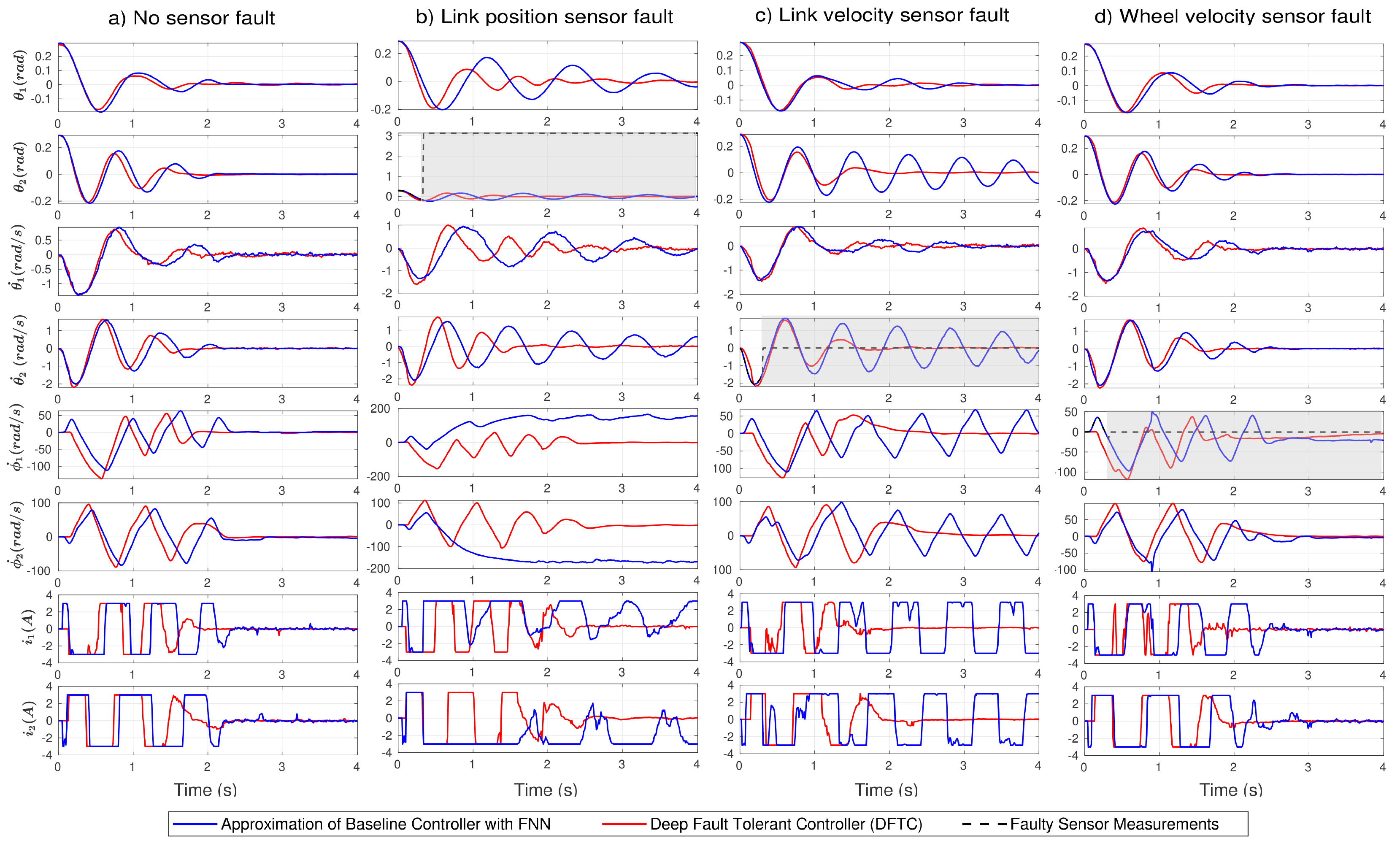}
        \caption{Experimental results of time evolution of states and inputs for FNN and DFTC with different sensor fault scenarios: a) No sensor fault, b) link position sensor fault, c) link velocity sensor fault, and d) wheel velocity sensor fault.}
        \label{fig:exp_results}
\end{figure*}

DFTC was deployed on the physical setup to validate its performance for real-time closed-loop control. Experiments with and without sensor faults were conducted similar to the simulation scenarios, but without the baseline controller (which cannot be implemented in real time). During the experiments, the inverted pendulum was released from approximately the same initial non-equilibrium point, resulting in an oscillatory motion of the system. Figure \ref{fig:exp_results} shows the experimental results of DFTC and FNN for different sensor fault scenarios. Also, the corresponding costs $J_{\rm DFTC}$ and $J_{\rm FNN}$ for these experiments are presented in Table \ref{tab:exp_cost}. 

\begin{table*}[t!]
    \centering
    \begin{tabular}{|c|c|c|c|c|c|c|}
    
    \hline
          & No sensor fault & Link position sensor fault & Link velocity sensor fault & Wheel velocity sensor fault \\
    \hline
         $J_{\rm DFTC}$ ($\times 10^{-4}$) & 30.12 & 30.80 & 30.39 & 31.80\\
    \hline
         $J_{\rm FNN}$ ($\times 10^{-4}$) & 35.60 & 72.61 & 46.53 & 34.88\\
    \hline
         $J_{\rm FNN}$ / $J_{\rm DFTC}$& 1.18 & 2.35 & 1.53 & 1.09 \\
    \hline
    
    \end{tabular}
    \caption{DFTC and FNN cost for experimental results.}
    \label{tab:exp_cost}
\end{table*}

\subsection{Discussion}
In the simulations, by analyzing Table~\ref{tab:rnn_cost}, one can see that DFTC and FNN performed similarly in case of no sensor fault; however, for those simulations containing abrupt sensors faults, the mean value of $\rho_{\rm DFTC}$ was less than half the mean value of $\rho_{\rm FNN}$. We also computed the average execution time for both DFTC and FNN control laws. Even though the DFTC computations took longer (around 1.4 ms, Lines 18-25 in Algorithm \ref{alg:dl_pipeline}), they were still within the $10$~ms sampling time, thus enabling real-time deployment in the experimental setup.

In the simulations with no sensor faults, both FNN and DFTC presented a performance similar to that of the baseline controller. This is shown in Fig. \ref{fig:sim_results}a, in which one can see that the state was steered to the origin after about $2.5$ s. On the other hand, when a fault happened in the link position or velocity sensor (Figs. \ref{fig:sim_results}b and \ref{fig:sim_results}c), the DFTC still performed well and stabilized the system with a small delay as compared to the baseline controller (the latter always considered in the absence of faults). On the other hand, the FNN controller could not stabilize the system, and the pendulum oscillated around the pivot point. Also, the wheel velocity increased to a large value resulting in significant energy expenditure. Lastly, with a wheel velocity sensor fault (Fig. \ref{fig:sim_results}d), both FNN and DFTC controllers stabilized the pendulum at its vertical position, but the angular velocity of the first reaction wheel using the FNN controller remained at a large value that decreased slowly to zero because of the system damping. 

% (Simulation) For the reaction wheel velocity sensor fault the story is:
% 1) both of the controllers stabilizes the system
% 2) both of them suppress the velocity of the wheel more or less instantly with no sensor fault
% 3) for the reaction wheel with faulty sensor, after the system stabilizes and all states become zero (with including the faulty wheel velocity), the model output (control input) is about zero which makes sense, and the reaction wheel with the faulty sensor comes to rest slowly with damping. 

% 4) but the difference shown in the Fig 4, last column: DFTC is more efficient  and the reaction wheel velocity does not reach high values. While FNN controller cause the large wheel velocity, which takes a while to become zero with damping. This is of course a bit extreme case for FNN
% 

The experimental results reported in Fig. \ref{fig:exp_results} are in line with the obtained simulation results. For no sensor fault, both controllers stabilized the system, with the cost $J_{\rm DFTC}$ significantly smaller (about 18\%) than the cost $J_{\rm FNN}$. This was probably due to the fact that DFTC might be a more robust controller due to the training, which includes abrupt sensor fault scenarios. Abrupt sensor faults can be regarded as discrepancies between the ideal model and the actual system responses; therefore, during experiments, DFTC might perform better, since it is inherently trained for a model-system parameter mismatch.

When the fault was introduced into link position or link velocity sensors, DFTC could compensate for the faulty sensor. On the contrary, FNN exhibited, as expected, an erroneous behavior: the control input to the system started to swing between the two extreme values, and the reaction wheels continued to run at high speeds. Also, the cost $J_{\rm FNN}$ presented an increase, as compared to $J_{\rm DFTC}$, of $135\%$ and $53\%$, respectively. Lastly, when the fault happened in the wheel velocity sensor, the FNN controller still stabilized the system, but small fluctuations in the control inputs resulted in non-zero wheel velocity. In contrast, using DFTC, $\dot{\phi}_1$ also converged to zero. In this case, the value of $J_{\rm FNN}$ was only $9\%$ higher than $J_{\rm DFTC}$.

It is hard to explore the underlying fault compensation mechanism because of the end-to-end nature of DFTC. With reference to Fig.~\ref{fig:lstm_feat}, the LSTM output features for the non-faulty states in both no sensor fault and link velocity fault cases were similar, while, for the faulty state, the LSTM output features were constant because of the deterministic features of the proposed DFTC model. However, the control system successfully stabilized the system. This suggests that the FC layers have some compensatory mechanisms for generating the required control actions in case the LSTM output for one sensor does not change by using the information from the other LSTM output features. In Fig.~\ref{fig:lstm_feat} we can also see that the time evolution of the states was approximately overlapping for the cases with and without fault. This means that, despite the fact that one sensor is faulty, the system still maintains a similar state trajectory as for the non-faulty case.

Our supervised learning approach for creating the DFTC utilized data augmentation using synthetic faults, and even the original data from the baseline controller were generated using simulations. One could foresee problems with real-world implementation because of the model-plant mismatch and naive assumption on the sensor fault effects. However, the experimental results highlight that simulations and synthetic data generation, in our case study, constituted a valid venue for control system development.

A video of the conducted experiments is provided as a supplementary material to the paper. The source codes and CAD models of the system are also uploaded to GitHub under MIT licence\footnote{\href{https://github.com/IS2AI/Deep_Fault_Tolerant_Control}{github.com/IS2AI/Deep\_Fault\_Tolerant\_Control}}.

\section{Conclusions and Outlook}\label{sec:concl}

This paper presented a data-driven approach to handle FTC problems using end-to-end neural networks. Extensive simulations and experiments on a mechatronic experimental setup showed that the presented DFTC scheme can succeed in stabilizing the system to the origin in the presence of abrupt sensor faults, without the need to separately design an FDI scheme and a control reconfiguration block.

The main limitations of the considered application case study are the following. Firstly, the design of the DFTC scheme for simultaneous faults of multiple sensors is not investigated. Secondly, the designed control scheme is not robust to changes in the system parameters such as mass, spring constant or the length of the link: if these parameters change, the design pipeline should be executed from its beginning to design a new RNN. Our future work will focus on investigating these issues (for example by using adaptive approaches, cf. \cite{yang2021new,yang2021adaptive}), and also on considering other neural network architectures, such as transformers.

\ifCLASSOPTIONcaptionsoff
  \newpage
\fi

%\balance

\bibliographystyle{IEEEtran}

\bibliography{biblio_file}

\begin{IEEEbiography}
[{\includegraphics[width=1in,height=1.25in,clip,keepaspectratio]{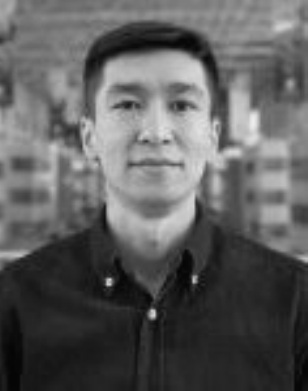}}]{Daulet Baimukashev}
received his B.S. and M.S. degrees in Robotics and Mechatronics from Nazarbayev University, Nur-Sultan, Kazakhstan in 2017 and 2019, respectively. During his M.S. studies he was a Research Assistant at the Advanced Robotics and Mechatronics Systems (ARMS) Laboratory, Nazarbayev University. Since 2019, he has been a Data Scientist at the Institute of Smart Systems and Artificial Intelligence, Nazarbayev University. His research interests include robotics, computer vision, and machine learning.
\end{IEEEbiography}

\begin{IEEEbiography}
[{\includegraphics[width=1in,height=1.25in,clip,keepaspectratio]{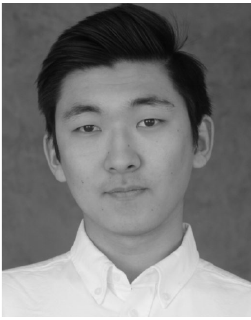}}]{Bexultan Rakhim}
received a B.S. degree in robotics and mechatronics engineering from Nazarbayev
University, Nur-Sultan, Kazakhstan, in 2018.
Since 2016, he has been a Research Assistant
with the Advanced Robotics and Mechatronics Systems (ARMS) Laboratory, Nazarbayev University.
His current research interests include machine learning, optimal control, and numerical methods in robotics.
\end{IEEEbiography}

\begin{IEEEbiography}
[{\includegraphics[width=1in,height=1.25in,clip,keepaspectratio]{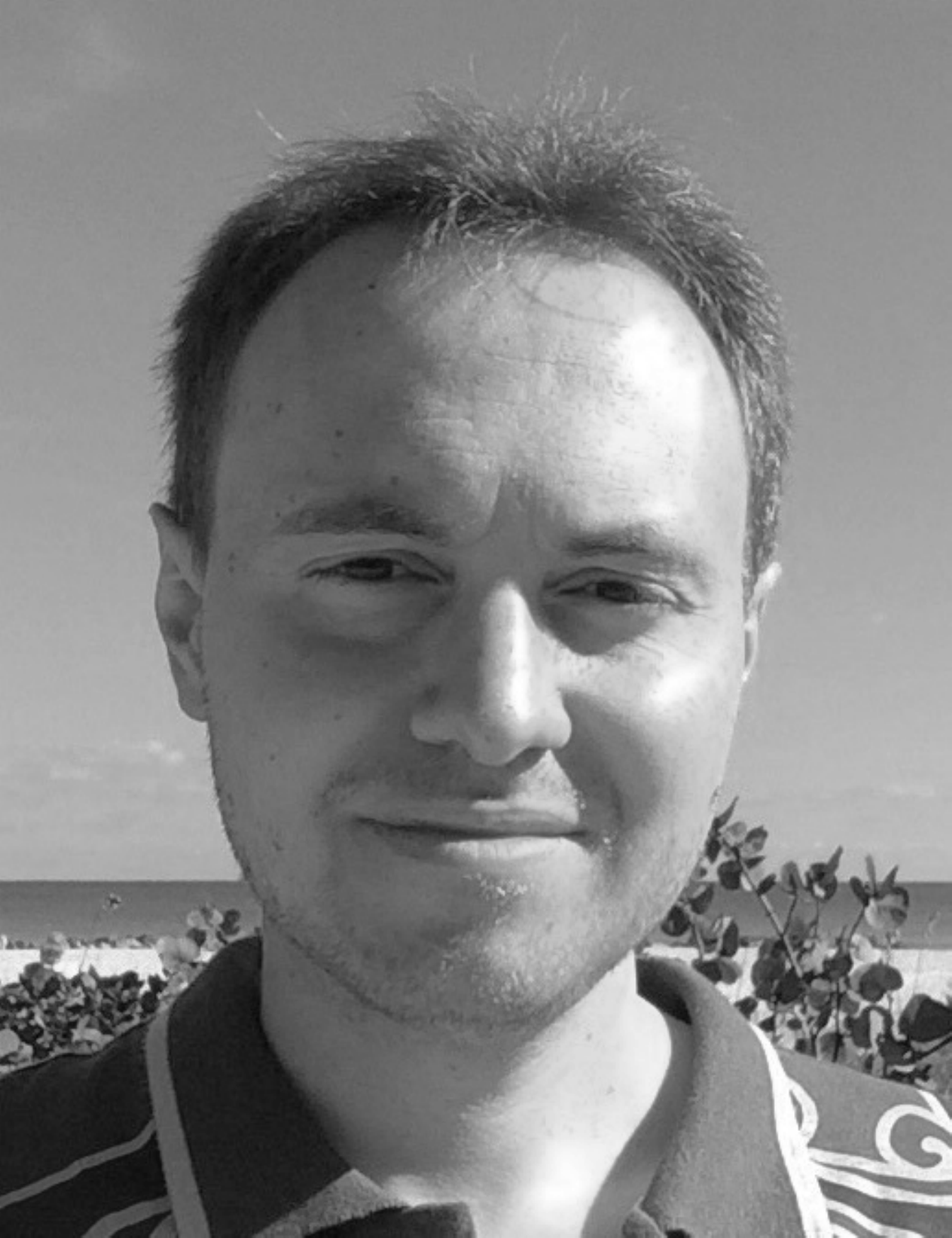}}]{Matteo Rubagotti}
(S'07, M'11, SM'19) received the Ph.D. degree in Electronics, Computer Science, and Electrical Engineering from the University of Pavia, Pavia, Italy, in 2010. Since 2018, he has been an Associate Professor of Robotics and Mechatronics at Nazarbayev University, Nur-Sultan, Kazakhstan. Previously to his post at Nazarbayev University, he was a Lecturer in Control Engineering at the University of Leicester, Leicester, UK, and a Postdoctoral Fellow at the University of Trento, Trento, Italy, and at IMT Institute for Advanced Studies, Lucca, Italy. He has co-authored more than 50 technical papers in international journals and conferences in the fields of control theory, robotics, mechatronics, and smart buildings. His current research interests include numerical optimal control, model predictive control and sliding mode control, and their applications to robotics and mechatronics.

Dr. Rubagotti currently serves as Subject Editor for the International Journal of Robust and Nonlinear Control, and is member of the conference editorial boards of the IEEE Control System Society and of the European Control Association.
\end{IEEEbiography}

\begin{IEEEbiography}[{\includegraphics[width=1in,height=1.25in,clip,keepaspectratio]{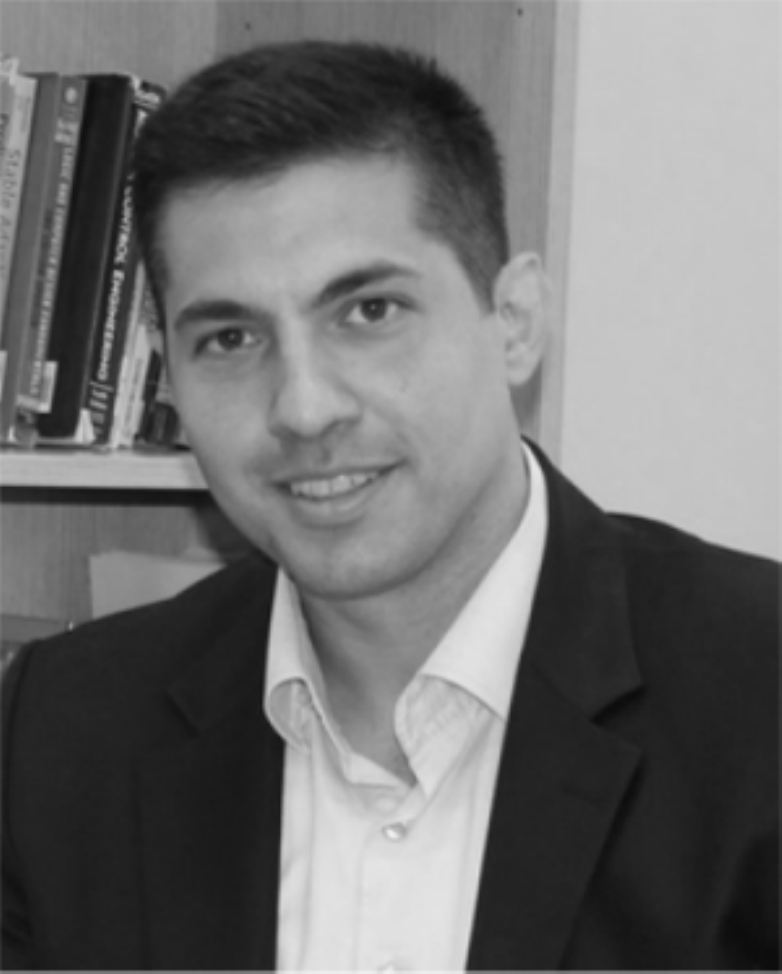}}]{Huseyin Atakan Varol}(M'09, SM'16) received a B.S. degree in mechatronics engineering from Sabanci University, Istanbul, Turkey, in 2005, and M.S. and Ph.D. degrees in electrical engineering from Vanderbilt University, Nashville, TN, USA, in 2007 and 2009, respectively. He was a Postdoctoral Research Associate and a Research Assistant Professor with the Center for Intelligent Mechatronics, Department of Mechanical Engineering, Vanderbilt University, from 2009 to 2011.
	
In 2011, he joined the Faculty of Nazarbayev University, Nur-Sultan, Kazakhstan, where he currently chairs the Department of Robotics and Mechatronics and directs the Institute of Smart Systems and Artificial Intelligence (ISSAI). He has published more than 60 technical articles on related topics in international journals and conferences. His research interests include biomechatronics, variable impedance actuation, machine learning, intelligent systems, and tensegrity.
	
Dr. Varol was a Finalist for the KUKA Innovation Award in 2014. He was also a recipient of the IEEE International Conference on Rehabilitation Robotics Best Paper Award in 2009 and the IEEE Engineering in Medicine and Biology Society Outstanding Paper Award in 2013. He serves as a Technical Editor of the IEEE/ASME Transactions on Mechatronics.
\end{IEEEbiography}

\balance
\end{document}